\documentclass[12pt,preprint]{aastex}

\usepackage{graphicx}

\slugcomment{Accepted to ApJ}

\shorttitle{Kinematic Center of NGC 253}
\shortauthors{M\"uller S\'anchez et al.}

\begin{document}


\title{The Stellar Kinematic Center and the True Galactic Nucleus of NGC253\footnotemark[1]
}


\author{F. M\"uller-S\'anchez$^2$, O. Gonz\'alez-Mart\'in$^{3,4}$, J.~A. Fern\'andez-Ontiveros$^{2,5}$, J.~A. Acosta-Pulido$^2$, M.~A. Prieto$^2$}

\affil{$^2$ Instituto de Astrof\'isica de Canarias, V\'ia L\'actea s/n, La Laguna, E-38205, Spain}

\affil{$^3$ Foundation for Research and Technology-Hellas, IESL, Voutes, 71110 Heraklion, Crete, Greece}

\affil{$^4$ University of Crete, Department of Physics, Voutes, 71003 Heraklion, Crete, Greece}

\affil{$^5$ Departamento de Astrof\'isica, Facultad de F\'isica, Universidad de la Laguna, Astrof\'isico Fco. S\'anchez s/n, La Laguna, E-38207, Spain}

\footnotetext[1]{Based on observations at the European Southern Observatory VLT (075.A-0250).}






\begin{abstract}
We present the first sub-arcsecond resolution two-dimensional stellar kinematics and X-ray observations of the prototypical starburst galaxy NGC253 which define the position and nature of the galactic nucleus. These observations comprise some of the best probes of the central 300 pc of NGC253, the nearest massive galaxy undergoing a powerful starburst, and will allow us to gain more insight into the nature of the centers of starburst galaxies.
We get an estimate of the stellar kinematic center location corresponding to an area of $r\sim 1.2^{\prime\prime}$ centered $\sim 0.7^{\prime\prime}$ southwest from the radio core, and historically presumed nucleus, TH2. 
Newly processed \textit{Chandra} data reveal a central point-like hard X-ray source (X-1) lying $\sim 0.4 \arcsec$ southwest from the kinematic center.
Very accurate alignment between radio, infrared and X-ray sources in the nuclear region shows that TH2, the IR photometric center and X-1 are not associated with each other. 
As the kinematic center is consistent with the positions of TH2 and X-1, and both could be a manifestation of nuclear activity, we consider the two as possible galactic nucleus candidates. 
Although TH2 is the strongest compact radio source in the nuclear region, 
it does not have any infrared, optical or X-ray counterparts. If the kinematic center is associated with this source, by analogy we suggest that the nucleus of NGC253 resembles our Galactic Center SgrA$^*$. 
On the other hand, X-1 is a heavily absorbed object ($N_{H} = 7.5 \times 10^{23}$ cm$^{-2}$) only detected at energies $>2$ keV ($L_{2-10 keV} \sim 10^{40}$ ergs s$^{-1}$). 
If X-1 is instead associated with the kinematic center, the nucleus of NGC253 is compatible with an obscured low luminosity active galactic nucleus (AGN) or a spatially resolved super star cluster (SSC) brightening up in X-rays most probably due to young supernovae or supernova remnants, a situation also observed in the nuclear starburst of M82. 
If no SSC is associated with the kinematic center, we conclude that NGC253 is a galaxy in which a strong starburst and a weak AGN (either TH2 or X-1) coexist. Results from few other high resolution studies of nearby starburst galaxies (e.g. M82, NGC5253, NGC4945) indicate that the AGN in these systems, if present, is always in the low luminosity regime.  
This may indicate that the onset of nuclear activity in galaxies is closely related with the occurrence of star formation, and that we are witnessing the emergence or dissaperance of an AGN. 

 
\end{abstract}

\keywords{galaxies: starburst --
 galaxies: individual (\objectname{NGC 253}) --
 galaxies: nuclei -- 
 galaxies: kinematics and dynamics-- 
 X-rays: galaxies --
 infrared: galaxies}


\section{Introduction} \label{int}


It is observationally well established that star formation in starburst galaxies manifests itself in the production of numerous young massive star clusters of extraordinary luminosity and compactness, normally referred to as super star clusters (SSCs). In fact, the four most luminous circumnuclear starburst systems (M82, M83, NGC253 and NGC4945) account for $25\%$ of the high-mass ($>8$ M$_\odot$) star formation within 10 Mpc \citep{heckman98}. Nearby starburst galaxies provide thus a unique laboratory for investigating the mechanisms which cause enhanced star formation to occur and be sustained, as well as physical conditions analogous to intensely star-forming galaxies identified at high-z (e.g. Shapley et al. 2003). 
An intriguing issue in studies of starburst galaxies is the apparent absence of a well defined nuclear source which could be associated with the center of the galaxy. 
This situation is present in spatially resolved starbursts in nearby galaxies: e.g. M82 \citep{matsumoto01}; NGC4945 \citep{marconi00}; NGC5253 \citep{summers04}; NGC253 (this work). 
Massive black holes (BHs) and/or nuclear star clusters (NSCs) are thought to reside in the centers of most galaxies, with the first being mostly identified in massive ellipticals \citep{richstone98, graham08} and NSCs in irregular and late spirals (e.g. B\"oker et al. 2002; Seth et al. 2010). Although tens of SSCs exist in the central region of starburst galaxies, the presence of a NSC in these systems is very poorly constrained. Since the central few hundred parsecs of starburst galaxies are heavily affected by large and nonuniform dust extinction, the determination of SSCs properties (age, colors, luminosities, etc.), and thus the detection of a putative NSC, has been difficult and uncertain. 
The presence of massive BHs in galaxy centers is normally inferred when nuclear activity is observed \citep{filippenko89}. 
Active Galactic Nuclei (AGN) are usually characterized by prominent point sources in $K$-band. Starburst galaxies do not show any IR point-like emission at the position of the assumed galactic nucleus, which in the majority of these systems has been associated to either a powerful compact radio source or a hard X-ray peak. However, in several cases it is not possible to reject/accept the AGN hypothesis on the basis of radio/X-ray emission peaks alone, particularly when the three photometric centers (radio, IR, X-ray) don't coincide. For the nearest, best studied starbursts, possible galactic center candidates include: young supernovae (SNe), supernova remnants (SNRs), intermediate-mass BHs, inactive supermassive BHs, low-luminosity AGN, etc. (see e.g. Wills et al. 1999 and Stevens et al. 1999 for the case of M82, Marconi et al. 2000 for NGC4945 and Fern\'andez-Ontiveros et al. 2009 for NGC253). 

As the center of a galaxy is likely to coincide with the bottom of the galaxy potential well, independent estimates of the kinematic center position from kinematic data can help to determine which photometric center, if any, is the most promising candidate for galactic nucleus. These important dynamical studies had not previously been widely performed because high S/N spectra over a full 2D field with high spatial resolution are required, and these are still scarce. 
In this paper we provide the first estimate of the kinematic center of the prototypical starburst galaxy NGC253 using sub-arcsecond resolution two-dimensional stellar kinematics. This is combined with data at radio, infrared and X-ray wavelengths to examine the nature of the galactic nucleus itself. These observations comprise some of the best probes of the inner region of NGC253 and will allow us to gain more insight into the nature of the centers of starburst galaxies.

As the nearest massive galaxy experiencing a strong starburst (3.94 Mpc, $1^{\prime\prime}$ = 19 pc, Karachentsev et al. 2003), 
NGC253 is one of the most studied objects in the sky,
yet many aspects of its nuclear region remain poorly understood. 
One of the main unresolved issues is the precise location and physical nature of the center of the galaxy. 
The galactic nucleus has historically been associated to a powerful compact non-thermal radio source 
(TH2 in Ulvestad \& Antonucci 1997) with characteristics similar to other synchrotron sources in active galaxies \citep{turner85}:
very small size ($<100$ mas),  
high brightness temperature ($T>40000$ K at 15 GHz), 
and a relatively flat spectral index between $8.4-23$ GHz($\alpha =-0.3$, $S_\nu \propto \nu^\alpha$). 
Follow up studies accross the electromagnetic spectrum, 
each limited by the spatial resolution available at that time, 
have matched this source with i) H$_2$O maser emission \citep{nakai95}, 
ii) different knots of IR emission \citep{forbes00, galliano05}, 
and iii) the hard X-ray emission peak \citep{weaver02}, 
suggesting that TH2 may be a low luminosity active galactic nucleus (LLAGN). 
However, it is now apparent from recent high-spatial resolution observations 
at different wavelengths, and using diverse observational techniques, that this picture may be incorrect. 
The evidences are the following:

\begin{itemize}
	\item VLA and VLBA observations at 23 GHz \citep{brunthaler09} 
have shown that the high velocity H$_2$O maser is not associated with TH2 but with another nearby continuum source (TH4 in Ulvestad \& Antonucci 1997), which has been classified as a SNR embedded in an HII region \citep{lenc06}. This suggests that a maser-AGN connection is unlikely and that the maser emission is almost surely related to star formation.

	\item Very high spatial resolution VLBI observations at 2.3 and 23 GHz \citep{lenc06, brunthaler09} did not detect TH2 on mas scales. As this source is unresolved and very bright in VLA 5-23 GHz continuum maps (beam sizes $>100$ mas), 
this questions the presence of a possible AGN. 

	\item Very accurate alignment at parsec scales has been found between the near-IR 
and radio sources by means of Adaptive Optics VLT/NACO IR data with a comparable resolution 
to those in existing VLA radio maps \citep{fernandez09}. 
Surprisingly, this new alignment shows that TH2 has no infrared counterpart. The newly revealed IR nature of TH2 shows no evidence of a central point source that would indicate nuclear activity. However, the non-detection of IR emission does not rule out other scenarios such as the existence of an inactive supermassive black hole (SMBH) like Sgr A$^*$ \citep{fernandez09, brunthaler09}. 
  
\end{itemize} 

In addition, one needs to be cautious when associating the hard X-ray peak to TH2. It is well known that NGC253 harbors $\sim 60$ compact radio sources \citep{ulvestad97} and $\sim37$ near-infrared point-like sources \citep{fernandez09}. This significant change in morphology with wavelength has led to different registrations in the literature (see e.g. Galliano et al. 2005 and references herein). Since \textit{Chandra} data were processed with $0.5\arcsec$ pixel $^{-1}$ resulting in $\sim 1 \arcsec$ resolution and have an astrometric accuracy of $\sim 1 \arcsec$ rms, the hard X-ray peak could be associated with several other discrete sources in the field. 
This is demonstrated in Figure 2 of \citet{weaver02} where a variety of IR and radio sources are contained within the size of the hard X-ray peak. 
Moreover, these authors followed the alignment proposed by \citet{forbes00} which, in light of the current high spatial resolution NACO images presented in \citet{fernandez09}, seems to be incorrect.  
Thus, it is still a topic of active investigation whether TH2 harbors an AGN, or even if it corresponds to the galactic center, or not.

Based on the alignment proposed by \citet{fernandez09} one could point to the infrared photometric center (or IR peak) as possible AGN candidate. The IR peak lies $\sim 3.3 \arcsec$ SW from TH2. It is in fact the brightest source at infrared wavelengths ($1-20\, \mu$m) and emits $\sim10\%$ of the galaxy's bolometric flux \citep{mohan02}. 
The spatial distribution of the 3.3 $\mu$m polycyclic aromatic hydrocarbons (PAH) feature-to-continuum ratio reaches a local minimum in this region \citep{tacconi05}. This indicates that the source produces strong ionization, probably destroying the PAH molecules by photo-dissociation. Supporting this interpretation is the fact that multiband tracers of ionized gas such as Pa$\beta$, Br$\gamma$, [HeI] and [NeII] are seen to peak strongly at this position \citep{boeker98, engelbracht98, keto99, forbes00, tacconi06}.
In addition, the IR peak has an individual counterpart at radio wavelengths (TH7 in Ulvestad \& Antonucci 1997) which also presents a relatively flat spectrum between $8.4-23$ GHz ($\alpha=-0.21$, $S_\nu \propto \nu^\alpha$), but it is $\sim 4$ times fainter than TH2. 
Moreover, this strong IR source is located at the vertex of the rather collimated ionised-gas region seen in NGC253 (see Figure 8 of Tacconi-Garman et al. 2005). All these characteristics make this object unique among the other sources in the central region of the galaxy and could be explained either by invoking a very young SSC or an AGN.

NGC253 has been the subject of many detailed kinematical
studies of its gaseous components. 
Optical rotation curves of the ionized gas have shown that, in addition to ordered rotation, large non-circular motions exist within the central region \citep{ulrich78, munoz93, arnaboldi95}. However, because of problems of obscuration
at optical wavelengths, results from H$\alpha$ and [N II]
observations are difficult to interpret. In addition, HI and H$\alpha$ rotation curves are incomplete: either they cover the outer parts $r>100\arcsec$ \citep{pence80}, or they are confined to the innermost regions where peculiar features and asymmetries in the ionized gas are dominant. 
Most of the
optical and IR observations, aimed at studying the kinematics
of the molecular gas in the nuclear region of NGC 253 (e.g., Prada et al. 1996; Engelbracht et al. 1998), have relied on measuring the velocity gradients along chosen position angles passing
through the nucleus (e.g., major axis and minor axis). 
Interferometric radio observations at millimeter and centimeter wavelengths measure two-dimensional velocity fields which provide a more complete picture of the kinematics in the nuclear region. 
Several radio observations in molecular emission lines
such as CO \citep{mauersberger96, canzian88, das01, paglione04}, CS \citep{peng96}, HCN \citep{paglione95}, or in
hydrogen recombination lines \citep{anantamariah96} exist in the literature. These two-dimensional measurements
reveal a complex but systematic velocity pattern in the
central region. In addition to solid-body rotation, the
observed velocity fields indicate motions which may be due
to a barlike potential in the center \citep{peng96} or a
kinematic subsystem which may be caused by a past merger
event (Anantharamaiah \& Goss 1996).
Although most of these observations
have high spectral resolution, they typically have
poor spatial resolution $>2.5 \arcsec$ and are not always useful for
probing the dynamics very close to the nucleus.

Because of this situation, different authors assumed different centers for their rotation curves/velocity maps: either the center is derived assuming complete symmetry of the outer parts in their fields of view, or it is coincident with the IR peak, or with the radio core TH2, or with the surface brightness peak. 
The highest resolution measurement of the two-dimensional
velocity field available to date is that of Anantharamaiah \& Goss (1996), who observed the H92$\alpha$ recombination line with a beam of $1.8\arcsec \times 1.0 \arcsec$ and a velocity resolution of 54 km s$^{-1}$. After trying several positions near the H92$\alpha$ emission peak, the optimum center of the velocity field was found to be at ($\alpha$, $\delta$)$_{2000}$ = $00^h47^m33^s.06$, $-25\degr 17^\prime 18\arcsec .3$ ($1\sigma$ error $\sim 0.3\arcsec$), which is $\sim 2.0 \arcsec$ southwest of the
H92$\alpha$ line peak which in turn is coincident with TH2. This means that the most accurate estimate of the kinematic center based on gas kinematics available to date lies closer to the IR peak ($\sim 1.3\arcsec$) than to TH2.

However, kinematic studies of galaxies focused on the gaseous component (ionized and/or molecular gas) have the potential for revealing the dynamical forces that work in the gas closer to the galactic nucleus, but they cannot trace adequately the gravitational potential. The gas represents only a fraction of the total mass of the galaxy and could be driven by non-gravitational perturbations such as interactions, warps, bars, etc. Therefore, the kinematic parameters of any nuclear galactic disk (center, position angle and inclination) can be greatly influenced by these perturbations. Thus the velocity field of stars is needed to properly trace the gravitational potential. 

Integral-field spectroscopy is an ideal tool for studying objects with complex morphologies and kinematics, such as the nuclei of starburst galaxies. 
This paper is the first of two describing the conclusions from an integral-field spectroscopic study with SINFONI on the VLT to clarify the structure, kinematics and emission mechanisms of the central 150 pc starburst region of NGC253. 
In this part we use the CO stellar absorption features in $K-$band to determine the stellar kinematic center of the galaxy. The stellar CO features have the advantage that they trace a component of the galaxy nucleus that is not affected by non-gravitational processes such as shocks or winds. 
We also obtained archival \textit{Chandra} data and re-processed them using $0.125\arcsec$ pixel$^{-1}$. The X-ray images are carefully compared with available high spatial resolution radio and near-IR data. This multiwavelength approach combined with the estimate of the kinematic center is essential to find the true nucleus of NGC253. 
In a forthcoming publication \citep{mueller09} we investigate the physical properties and kinematics of the super star clusters and the molecular and ionized gas in the nuclear starburst. 

This paper is structured as follows. In Section 2, we describe the SINFONI and \textit{Chandra} observations and the processing we performed on the data. The method used to extract the kinematics from the SINFONI spectral features is presented in this Section as well as the new X-ray images and spectra. In Section 3 we examine the stellar velocity field and obtain an estimate of the position of the kinematic center. We discuss the implications of our results for the nature of the galactic nucleus in Section 4. The overall conclusions of the study are outlined in Section 5.   

\section{Data Analysis and Results} \label{obs}

\subsection{Near-IR SINFONI data}\label{sinfoni}

\subsubsection{Observations and data reduction}\label{sinf_reduc}

The SINFONI instrument on the VLT \citep{bonnet04, eisenhau03} was used without the adaptive optics module, resulting in slice widths of $0.25\arcsec$, a pixel scale of $0.125\arcsec$ pixel$^{-1}$, and a field of view of $8\arcsec \times 8 \arcsec$. 
Due to the high-inclination of the galaxy ($\sim 78\degr$), this field size is not enough to cover the diameter of the circumnuclear starburst regions under study, hence mosaicing was necessary. To cover the central $30\arcsec$ along the photometric major axis, 8 different pointings were performed and the raw data were obtained from the ESO archive. The individual frames were combined to form a mosaic that was used for analysis. The SINFONI mosaic has a dimension of $30\arcsec \times 30\arcsec$ and is centered at the position of the presumed galactic nucleus, TH2.

The SINFONI data presented here were taken in $K$-band (1.9--2.4 $\mu$m) on the night of 28 August 2005 in 18 individual frames of 300s resulting in 1.5h total integration time. In this night the atmospheric conditions were excellent (seeing $< 0.7\arcsec$), which allows us to obtain a resolution of 0.5$\arcsec$ as measured from the FWHM of the smallest knot of emission in $K$-band. The FWHM spectral resolution is 68 km s$^{-1}$ (R$\sim$4400). The data were reduced using the SINFONI custom reduction package SPRED \citep{abuter05}. This performs all the usual steps needed to reduce near infrared spectra, but with the additional routines for reconstructing the data cube. Flux calibration was performed using a standard star observed next in time to the science frames, 
which yielded in a 0.5$\arcsec$ aperture centered at the near-IR emission peak a $K$-band magnitude of 12.34. 
In addition, flux calibration was cross-checked with NACO data in a similar aperture \citep{fernandez09}. Agreement between the different data sources was consistent to 15\%.

\subsubsection{Spectral fitting}\label{starfit}

The shape of the CO absorption bandheads in a galaxy spectrum arises from the intrinsic spectral profile of the stars themselves convolved with a broadening profile which carries the information about the kinematics. The use of template stars 
allows one to separate these two components.
We have extracted the 2D properties of the CO absorption features by  
fitting a Gaussian convolved with a template stellar spectrum to the
continuum-subtracted spectral profile of the CO bandheads at each spatial pixel in the data cube. 
For the CO bandheads in NGC253, we chose a template stellar spectrum of a late spectral type star, HD49105 (spectral type K0), observed with SINFONI in the same instrument configuration (pixel scale and grating). 
We performed a minimisation of the $\chi^2$ in which the parameters of the Gaussian (amplitude, center and width in velocity space) were adjusted until the convolved profile best matched the data ($\chi^2/$dof $\approx1$). The uncertainties were boot-strapped using
Monte Carlo techniques, assuming that the noise is uncorrelated and the intrinsic profile is well represented by a Gaussian \citep{davies07}. 
This method allowed us to obtain uncertainties for the velocity map in the range of $(5-15)$ km s$^{-1}$.  

 
Thanks to the high S/N of our data we could make use of the three overtone bands of $^{12}$CO at 2.294 (2-0), 2.323 (3-1), and 2.353 (4-2) $\mu$m. This is demonstrated in Fig.~\ref{fig1}, where we have plotted the spectral region around the $K-$band CO bandheads of several knots of IR emission in the nuclear region. 
We have extracted stellar flux, velocity and dispersion maps. However, in this paper we only analyze the stellar velocity field and refer to \citet{mueller09} for a detailed study of the other maps. 

\subsubsection{Distribution of the $K-$band continuum emission}\label{morphology}

\begin{figure}
\epsscale{.99}
\plotone{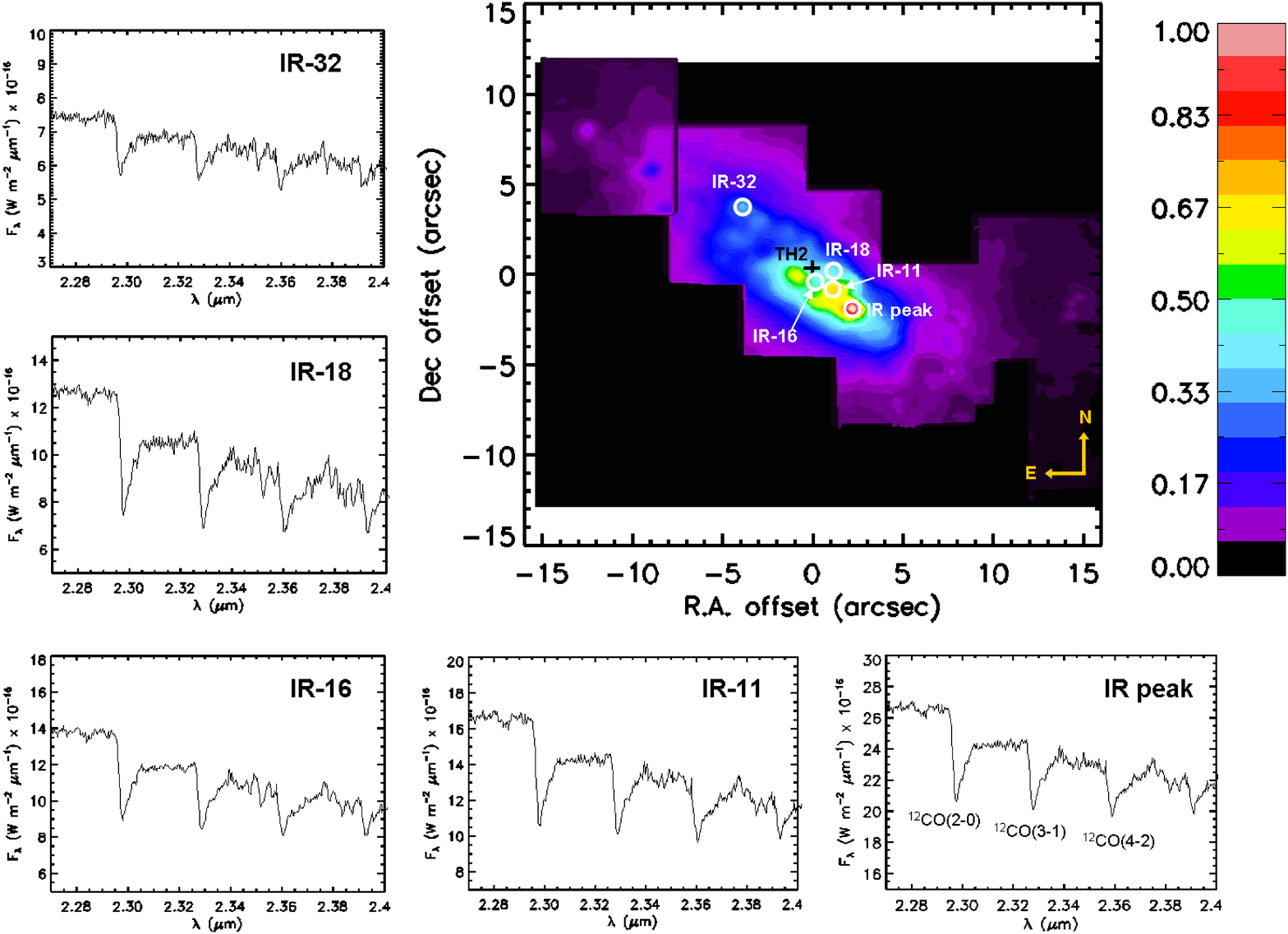}
\caption{SINFONI $K-$band continuum emission map of the central $30\arcsec \times 30\arcsec$ ($570\times 570$ pc) of NGC253. 
The image is centered at the position of the radio core, and presumed galactic nucleus, TH2 (black cross). This corresponds to an absolute position of 
($\alpha$, $\delta$)$_{2000}$ = $00^h47^m33^s.17$, $-25\degr 17^\prime 17\arcsec .1$. Figures 1, 4 and 5 (all SINFONI maps) are centered at this position. 
Relevant IR sources are labeled as ``IR-$n$'' following \citet{fernandez09} enumeration, except for the IR peak. The spectra of the five IR sources indicated in the Figure in the wavelength range between 2.27 and 2.4 $\mu$m are shown. The three overtone bands of $^{12}$CO at 2.294 (2-0), 2.323 (3-1), and 2.353 (4-2) $\mu$m, used for the extraction of the stellar kinematics are clearly seen. The spectra were extracted within a circular aperture of $r=0.25\arcsec$.
\label{fig1}}
\end{figure}

A flux map of $K$-band continuum emission in the central $570 \times 570$ pc of NGC253 is presented in Fig.~\ref{fig1}. The image is centered at the 
position of the radio core, and presumed galactic center, TH2 \citep{ulvestad97}. 
Registration of sources is based on the alignment between radio and near-IR images proposed by \citet{fernandez09}. Relevant IR sources in this study are labeled as ``IR-$n$'' following the enumeration used by these authors as well, except for the IR peak. 
Here we briefly summarize the main morphological features of the $K-$band continuum to put them in context for the upcoming kinematic analysis and discussion. 

Our SINFONI $K-$band map agrees well with previous near-IR studies of slightly lower spatial resolution \citep{sams94, engelbracht98, forbes00} and with the high-resolution Adaptive Optics VLT/NACO images presented in \citet{fernandez09}. 
At these wavelengths (1.9-2.4 $\mu$m) the continuum emission of the nuclear region consists of multiple compact sources distributed along the disc of the galaxy in an apparent ring-like structure superposed on a diffuse emission component. 
The spatial resolution of our data and the availability of mosaic images  
at $H-$ and $K-$bands 
providing a larger field of view compared to standard SINFONI observations, allow us to identify several discrete sources in the central $300 \times 300$ pc of the galaxy. 
A qualitative and quantitative analysis of the intrinsic properties of the stellar clusters found in the central 300 pc of NGC253 will be presented in \citet{mueller09}.  

As can be seen in Fig.~\ref{fig1}, the IR photometric center (the point of maximum emission in the wavelength range 1 -- 20 $\mu$m) is offset by about $\sim 3.3\arcsec$ (60 pc) from the radio core, TH2. While the IR peak has an individual counterpart at radio wavelengths (TH7 in Ulvestad \& Antonucci 1997), there is no discrete IR source associated with the radio nucleus \citep{fernandez09}. 
As it was discussed in the Introduction, this situation has originated a profuse debate about the precise location of the center of the galaxy and the nature of these two photometric centers. 
Thanks to the availability of sub-arcsecond resolution integral field data we are able to provide a third independent estimate of the center of the galaxy: the stellar kinematic center. This will be derived in Sect.~\ref{kinemetry} and compared with radio, IR and X-ray images of the nuclear region in Sect.~\ref{align}.

\subsection{X-ray \textit{Chandra} Data}\label{chandra}

\subsubsection{Observations and Data Processing} \label{chandra_reduc}

Level 2 event data from the ACIS instrument were extracted from
\emph{Chandra} archive\footnote{http://cda.harvard.edu/chaser/} (ObsID 3931) and observed in 2003 September 19. 
The data were reduced with the {\sc ciao 4.0}\footnote{See
http://asc.harvard.edu/ciao} data analysis system and the 
\emph{Chandra}  Calibration Database (caldb 4.0.0\footnote{see http://cxc.harvard.edu/caldb/}). 
The exposure time was processed to exclude background flares, using the task {\sc
lc\_clean.sl}\footnote{see http://cxc.harvard.edu/ciao/download/scripts/} 
in source-free sky regions of the same observation. The net exposure time 
after flare removal is 76~ksec. The nucleus has no significant pileup.

\emph{Chandra} data include information about the photon energies 
and positions that were used to obtain energy-filtered images, and 
to carry out sub-pixel resolution spatial analysis. Although the default
pixel size of the \emph{Chandra}/ACIS detector is $\sf{0.492\arcsec}$, smaller 
spatial scales are accessible as the image moves across the detector
pixels during the telescope dither, therefore sampling pixel scales smaller than 
the default pixel of {\it Chandra}/ACIS detector. It allows sub-pixel binning of the images.
We made 0.125$\sf{\arcsec~pixel^{-1}}$ images with FWHM resolution of $\sim 0.5\arcsec$ in two bands: 0.5-2 keV (hereafter soft band) and 
2-10 keV (hereafter hard band). The absolute astrometric accuracy of \emph{Chandra} data is
1$\sf{\arcsec}$ rms.

In addition to the high-spatial resolution analysis we applied 
smoothing techniques to detect the low contrast diffuse emission.  
We applied the adaptive smoothing CIAO tool {\sc csmooth}, based on the algorithm
developed by \citet{ebeling06}. 
{\sc csmooth} is an adaptive smoothing tool for images containing 
multiscale complex structures, and it preserves the spatial signatures
and the associated counts as well as significance estimates.
A minimum and maximum significance S/N level of 3 and 4, 
and a scale maximum of 2 pixels were used. 

The central sources detected in the image analysis were extracted to make 
the spectral analysis using circular apertures centered in the sources and with small radii of $r\sim0.5\arcsec$. 
This radius was chosen as trade-off between number of counts and contamination from background emission. We then extracted the background from source-free apertures around the hard X-ray emission regions. We subtracted the background from each source and performed the spectral analysis in the background-free sources. 
Response and 
ancillary response files were created using the CIAO {\sc mkacisrmf} and 
{\sc mkwarf} tools. To be able to use the $\sf{\chi^{2}}$ as the fit 
statistics, the spectra were binned to give a minimum of 20 counts per 
bin before background subtraction using {\sc grppha} task, included 
in FTOOLS\footnote{http://heasarc.gsfc.nasa.gov/docs/software/ftools}.
The analysis of the spectral counts was performed using 
the software package XSPEC (version 
12.4.0\footnote{http://cxc.heasarc.gsfc.nasa.gov/docs/xanadu/xspec/}, 
Arnaud et al. 1996).


\subsubsection{X-ray imaging}\label{ximages}

\begin{figure}
\epsscale{.99}
\plotone{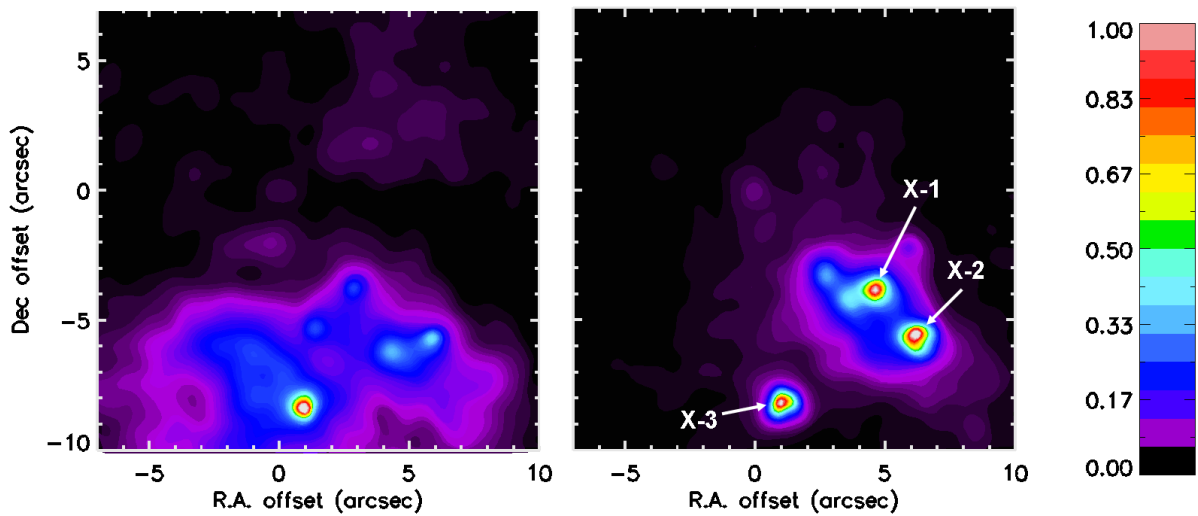}
\caption{\textit{Chandra} soft (0.5-2 keV, left) and hard (2-10 keV, right) X-ray images of the central $17\arcsec \times 17\arcsec$ region of NGC253. In the two panels north is up and east is left. The images are centered at the position of source IR-32, unambiguously identified at infrared, optical and X-ray wavelengths. This corresponds to an absolute position of 
($\alpha$, $\delta$)$_{2000}$ = $00^h47^m33^s.4$, $-25\degr 17^\prime 14\arcsec .0$. Figures 2, 6 and 7 (all \textit{Chandra} images) are centered at this position. 
The locations of the three hard X-ray sources discussed in the paper are marked in the right panel following the nomenclature ``X-$n$''.\label{fig2}}
\end{figure}

Fig.~\ref{fig2} shows the newly processed $\sim 0.5\arcsec$ resolution ACIS images of the nuclear region ($17\arcsec \times 17\arcsec$) of NGC253 in two energy bands: soft (0.5-2 keV) and hard (2-10 keV) X-rays. 
The images are centered at the position of source IR-32, unambiguously identified at infrared, optical and X-ray spectral ranges (see Sect.~\ref{align}). 

The soft X-ray image shows several knots of emission embedded in a diffuse component which originates in the nuclear starburst and extends southwest in an apparent cone of ionization. Indeed, \citet{strickland00} showed that the soft X-ray emitting gas distribution is consistent with a conical outflow of hot plasma, that is well collimated in a direction nearly perpendicular to the plane of the galaxy and extremely similar in morphology and extent to the previously
observed limb-brightened H$\alpha$ outflow cone \citep{lehnert96}.


In the hard X-ray image the emission is more compact and less diffuse. Three individual point-like sources are clearly identified: two lying in the nuclear starburst (X-1 and X-2) and one located $\sim 6 \arcsec$ SE from the nuclear region which corresponds to X-3 (Fig.~\ref{fig2}).
We use the ``X-$n$'' nomenclature because it denotes easily the number of individual point-like hard X-ray sources found in the field (Fig.~\ref{fig2}). X-1 and X-2 are separated by $\sim2.3 \arcsec$ (44 pc). While X-2 has a counterpart at soft X-rays, X-1 is only bright at energies $>2$ keV, suggesting that this source is highly absorbed. 
Extracting an even harder image in the 4--10 keV band shows that indeed X-1 is the brightest point-like source at higher energies. This is consistent with the spectra of the three sources (see Sect.~\ref{xspectra} and Fig.~\ref{fig3}).


\subsubsection{X-ray spectroscopy}\label{xspectra}

We performed spectral analysis for the three hard X-ray sources identified in the right panel of Fig.~\ref{fig2}. We investigated five baseline models for the three sources, two of them only with one component: abs(power-law) and abs(MEKAL); and the other three composite models: abs(power-law)+abs(MEKAL), abs(power-law)+abs(power-law) and abs(power-law+MEKAL)+abs(power-law). The best fit was selected as the simplest model in which we can reproduce the observed spectrum based on $\chi^2$ statistics. We tested whether a more complex model is statisticaly justified by using f-statistic ({\sc ftest} within Xspec). See \citet{gonzalez09} for a detailed description of the models and the fitting process. 
The spectra of the three sources are shown in Fig.~\ref{fig3} and the model parameters are listed in Table~\ref{table1}.  


\section{Stellar kinematics}\label{stars}

\subsection{Evidence for a rotating stellar disc}\label{kinematics}

We have extracted stellar kinematics in the nuclear region of the galaxy using the method described in Sect.~\ref{starfit}, allowing for the first time a two-dimensional study of the stellar motions on scales enclosing the star-forming nuclear ring. Fig.~\ref{fig4} shows the velocity field of the stars in the central $30\arcsec \times 30\arcsec$ of the galaxy. 
As can be seen in this Figure, the stellar velocity field is very regular, typical of rotational motions. The major kinematic axis exhibits a velocity gradient from -80 km s$^{-1}$ in the NE to 80 km s$^{-1}$ in the SW and is well aligned with the photometric major axis of the nuclear disc. In addition, it presents azimuthal symmetry and the zero-velocity axis (or kinematic minor axis) is perpendicular to the kinematic major axis, as expected in regions dominated by rotation. 
The observed regular pattern of the stellar velocity field permits us to use a simple axisymmetric disc model to characterize the stellar kinematics in the nuclear region, and particularly, to constrain the location of the kinematic center. 
The fitting method and the derived model parameters will be discussed in the following Section.   



\subsection{Kinematic modeling: Defining the kinematic center of the galaxy}\label{kinemetry}

If a velocity map exhibits ordered rotation, such as the stellar velocity field of NGC253, the velocity can be well represented by circular motion in a disc. 
The standard method for deprojecting galaxies from
kinematic data is to assume that the galaxy is made up of a
number of rotating coplanar annuli. 
The velocity field is fitted with the equation: 
\begin{equation}
V(r,\phi) = V_{sys} + V_c(r) \cos(\phi - \alpha) \sin i \nonumber
\end{equation}
where $r$ is the radius of a circular ring in the plane of the galaxy (or the semimajor axis of the ellipse in the plane of the sky), $V_{sys}$ is the systemic velocity, $V_c$ is the ring circular velocity, $i$ is the ring inclination, $\alpha$ is the position angle of the projected major axis, and $\phi$ is the azimuthal angle. This is the cosine law approximation \citep{krajnovic06}. 
A crucial element of our investigation, which is intrinsic to this model, involves choice of the kinematic center. Most kinematic studies of galaxies do not consider the center as a variable, as they know a priori its location.
For the purpose of this study, the stellar velocity field shall be fitted with a single disc model with four free geometric parameters: the center $(x_0, y_0)$, the position angle $\alpha$, and the inclination $i$.


The parameters of the rotating disc can be well determined via the kinemetry method developed and described in detail by \citet{krajnovic06}. Briefly, kinemetry is an extension of surface photometry to the higher-order moments of the velocity distribution. The procedure operates by first describing the data by a series of concentric ellipses of increasing major axis length, as defined by the system center, position angle, and inclination. Normally, the location of the center is the only parameter that is determined prior to running a
kinemetric analysis; the latter two parameters (position angle and inclination) can either be determined a priori and used as inputs or be measured functions of semi-major axis length. At each radius, a small number of harmonic terms in a Fourier expansion is needed to determine the best-fitting ellipse:  
\begin{equation}
K(\phi) = A_0 + \sum^N_{n=1} A_n \sin(n\phi) + B_n \cos(n\phi) \nonumber
\end{equation}
where $K$ is the best-fitting ellipse at a given radius, $N$ is the maximum fit order used, and $\phi$ is the eccentric anomaly, which in the case of discs corresponds to the azimuthal angle. Following \citet{schoenmakers99}, a decomposition of the velocity fields up to third order (i.e., $N = 3$) is sufficient to capture most of the signal. 
If the motion is purely circular, the observed velocity along that
ellipse will be described by a cosine law. In this ideal case, the $B_1$ term should be the only coefficient present in the harmonic expansion.  
Power in the other coefficients represents either deviations from circular motion (bars, inflows/outflows, etc.), or artifacts induced by the noise or incorrect geometric parameters (kinematic center, position angle and inclination). 

Our fitting procedure therefore consists of iteratively changing the four geometric parameters ($x_0$, $y_0$, $\alpha$, $i$) until the values of $A_0$, $A_1$, $A_2$, $A_3$, $B_2$, and $B_3$ are zero or close to zero. In other words, we use the kinemetry software to construct the best fitting rotating disk model of the stellar velocity field and determine simultaneously the kinematic center, position angle and inclination. In contrast to other previous kinemetric studies \citep{krajnovic07, boeker08, shapiro08}, in which the center is known a priori and the position angle and inclination are determined in two steps, the nature of our problem suggests that we need to calculate the four parameters simultaneously, as the location of the kinematic center is thigthly connected with the position angle of the kinematic major axis. Note that this procedure affects the determination of non-circular motions and could lead to slightly different geometric parameters $\alpha$ and $i$ as for the cases in which the centers are known a priori. 
The position of the kinematic center is to first order that
position which minimizes the non-circular motions in a tilted-ring fit and adopting this center position would then result in non-circular motions slightly smaller than the ones derived for the cases of known kinematic centers. 

The regular pattern of the stellar velocities allows us to consider the entire velocity field at once and solve for a global kinematic center, position angle and inclination. In fact, we do not consider the possibility of having a kinematic center which varies with radius as it does not have a physical meaning. 
In more detail we performed the following steps:

\begin{enumerate}
	\item We masked out pixels in the velocity field having $\le 10\%$ of the peak flux in the $K-$band continuum and those presenting velocities greater than $\left| 150 \right|$ km s$^{-1}$. 
	\item We assumed initial values of ($x_0$, $y_0$, $\alpha$, $i$) for the stellar disc in units of pixels for ($x_0$, $y_0$) and degrees for ($\alpha$, $i$), and kept them constant through the kinemetric analysis. The choice of these units was dictated by practical reasons of easily identifying the kinematic center in our SINFONI maps. 
	\item The observed velocity field $V$ was sampled along a set of ellipses regularly spaced in the semimajor axis and with constant ($x_0$, $y_0$, $\alpha$, $i$) as defined above. 
	\item The velocities of every ellipse were fitted with a complete kinemetric expansion.
	\item We determined the agreement between the observed velocity
field $V$ and the disc model $V_m$ (the 2D image reconstructed
from only the $B_1$ term of the kinemetric expansion) as: 
$\chi^2 = \sum^N_j[(V_{m,j} - V_j)/\Delta V_j]^2$
where $N$ is the number of pixels included in the fit and $\Delta V_j$ are the measured errors of the velocity at each pixel $j$. The residuals map $(V_m - V)$ reflects the combined powers in the higher coefficients, which may appear as asymmetries in the velocity field. As our stellar velocity field shows ordered rotation, the residuals are a measurement of the error in the estimation of the disc parameters.
	\item We then performed a multidimensional minimization of the function $\chi^2$($x_0$, $y_0$, PA, $i$) using the AMOEBA function in IDL \citep{nelder65}.
	\item We finally selected the best fitting values and the
corresponding errors from the parameter space of $\chi^2$($x_0$, $y_0$, $\alpha$, $i$). The values of ($x_0$, $y_0$, $\alpha$, $i$) that give the lowest $\chi^2$, ${\chi^2}_{min}$, were considered to be the best-fit values and the confidence intervals were determined by a range of $\chi^2$ values greater than ${\chi^2}_{min}$, defined as
$\Delta \chi^2 = \chi^2 - {\chi^2}_{min}$. If it is assumed that the observational errors are normally distributed, then $\Delta \chi^2$ also follows a $\chi^2$ probability distribution with the number of degrees
of freedom equal to the number of model parameters
\citep{lampton76, vandermarel98}. 
\end{enumerate}

The result of the above procedure provided the best fitting
disc parameters acceptable to the $99.7\%$ confidence level (corresponding to a $3\sigma$ uncertainty): the center is located at pixel $x_0 = 129^{+11}_{-8}$, $y_0 = 104^{+8}_{-10}$ in our data, $\alpha = 50 \pm 11 \degr$, and $i = 77^{+5}_{-13}$ degrees. 
For reference, TH2 is located at pixel $x=124$ and $y=107$ in our SINFONI maps, which implies that the kinematic center is offset by $\Delta x = 0.6\arcsec$ and $\Delta y = -0.4\arcsec$ from the radio core. Notice that the position of TH2 is consistent with the kinematic center estimate within the errors. 
The estimated absolute position of the kinematic center is thus ($\alpha$, $\delta$)$_{2000}$ = $00^h47^m33^s.14$, $-25\degr 17^\prime 17\arcsec .52$ 
with a mean $3\sigma$ uncertainty of $r\sim1.2\arcsec$. 
The stellar velocity field used for the modeling is presented in the left panel of Fig.~\ref{fig5}, 
while the best fitting disc model is shown in the middle panel.
The residuals between the data and the best fitting model are
displayed in the right panel. 
It is apparent from the residuals map that the observed velocities can be remarkably well described by the adopted simple thin disc approximation. 
No obvious regular patterns are noticeable in the residuals map, suggesting that the features observed in the residuals are only due to noisy pixels in the data. Furthermore, this map shows values between $[-15, 15]$ km s$^{-1}$ consistent with the measured errors of the velocity map. The best fit $\alpha$ and $i$ are in agreement with those from the photometric nuclear ring and those of the galactic disc \citep{pence80, paglione04}. 
Regarding to the location of the stellar kinematic center, we found that it lies in between the two original candidates for the galactic center, TH2 and the IR peak. However, the estimated confidence interval of our kinematic modeling clearly rules out the IR peak as kinematic center
(they are separated by $\sim 2.6 \arcsec$ distance from each other). A more detailed discussion on the candidates for galactic nucleus and their physical properties will be presented in Sect.~\ref{candidates}. 




\section{The Nucleus of NGC253}\label{candidates}

\subsection{Alignment between radio, IR and X-ray images of the nuclear region} \label{align} 

Fig.~\ref{fig6} summarizes the main results of our investigation. In this Figure, an overlay of the continuum emission at radio, IR and hard X-ray wavelengths in the central $10\arcsec \times 10\arcsec$ ($\sim 200 \times 200$ pc) of the galaxy is shown. 
The colour scale corresponds to the high resolution 3.8$\mu$m NACO image from \citet{fernandez09}. The 2-cm VLA A-configuration map from \citet{turner85} is shown in white contours and the new 2-10 keV \textit{Chandra} image from this work is presented in blue contours. The relevant radio, infrared and X-ray sources are indicated in the Figure, and the stellar kinematic center within approximately the $3\sigma$ errors is marked with a crossed circle of radius $r=1.2 \arcsec$. We use the nomenclature from \citet{ulvestad97} for the radio sources (except for source 5.73--39.5), and the IR and hard X-ray sources are labeled following the nomenclatures proposed in Sect.~\ref{morphology} and Sect.~\ref{ximages}, respectively. The alignment between radio and near-IR images was done as proposed by \citet{fernandez09}. The key alignment, however, is between hard X-ray and IR images.
The first registration of sources was obtained directly from \textit{Chandra's} absolute astrometry. The absolute astrometric accuracy of the nuclear pointing is $\sim 1 \arcsec$ rms. This astrometric approximation could be improved 
through the identification of common sources. We noticed that two of the brightest IR point-like sources in the nuclear region, the IR peak and IR-32, 
have individual counterparts in the X-rays (see Fig.~\ref{fig6}). 
The presence of these reference objects allowed us to tie the X-ray and IR images and achieve $\sim 0.5\arcsec$ relative position accuracy in searching for radio and IR counterparts to the X-ray sources. 
The alignment was further cross-checked 
using optical and X-ray images. A three-color composite $HST/Chandra$ image of the central $20\arcsec \times 20\arcsec$ ($400 \times 400$ pc) region of NGC253 is shown in Fig.~\ref{fig7}. Green corresponds to $V$-band (WFPC2/F555W), red to soft X-ray and blue to hard X-ray emission. The image is centered at the position of source IR-32, which is an outstanding feature at IR, optical and X-ray wavelengths, and a reference object in the alignment. We found that the extended soft X-ray emission southeast of the nuclear starburst is well correlated with the $V$-band emission in this region. Furthermore, the most prominent dark dust lane running along the northwestern part of the nuclear starburst is spatially coincident in optical and soft X-ray images (see Fig.~\ref{fig7}). 


As can be seen in Fig.~\ref{fig6}, within the $3\sigma$ error circle of the stellar kinematic center, there are $\sim 15$ objects detected at different wavelengths: 6 radio sources, 8 IR knots of emission interpreted as young star clusters \citep{fernandez09}, and the hard X-ray source X-1. 
Having ruled out the IR peak as the kinematic center, the two next likely candidates for galactic nucleus are the photometric centers at radio and X-ray wavelengths (TH2 and X-1, respectively), the two consistent with the kinematic center location. 
Surprisingly, TH2 and X-1 are not each other's counterparts. X-1 is located $\sim1.1 \arcsec$ SW from TH2 well beyond the error of $\sim0.5 \arcsec$ in the registration of the images. 
Not detecting any point-like hard X-ray source at the position of TH2 questions strongly its AGN nature, but does not discard it as the galactic nucleus with characteristics similar to Sgr\,A$^*$, as proposed by \citet{fernandez09}. 
However, \textsl{Chandra} results clearly point to X-1 as AGN candidate.
As our analysis cannot conclude whether the stellar kinematic center corresponds to TH2 or X-1, we consider the two as possible galactic nucleus candidates and discuss the implications in the following subsections. 

As X-2 lies $\sim 2.6 \arcsec$ (50 pc) SW from the kinematic center and X-3 is located $\sim 6 \arcsec$ (120 pc) away from the nuclear region (see Fig.~\ref{fig6}), the discussion of their properties is out of the scope of this paper. However, they are still presented 
in Fig.~\ref{fig3} and Table~\ref{table1} for completeness.


\subsection{TH2 as candidate for galactic nucleus}\label{th2}

The newly processed \textit{Chandra} observations of NGC253, 
combined with most of the current high-spatial resolution studies of this galaxy at different wavelengths, provide sufficient evidence against the AGN nature of TH2 (see Sect.~\ref{int}).
In fact, this source is only detected at radio wavelengths between 2.3 and 23 GHz and using baselines corresponding to synthesized beam sizes $>30$ mas \citep{sadler95, lenc06}. At frequencies lower than 2.3 GHz the source is not visible probably due to strong free-free absorption \citep{tingay04}. 
Longer baselines completely resolve TH2 as demonstrated by recent VLBI observations at 2.3 and 23 GHz with $15 \times 11$ and $2.5 \times 1.1$ mas resolution, respectively \citep{lenc06, brunthaler09}. The upper limit on the flux density of these two observations is $\sim 1.1$ mJy. As this source is unresolved with the VLA (size $<100$ mas and a flux density of 20 mJy at 23 GHz), this implies that its linear size is probably in the range of $0.5-2$ pc. \citet{lenc06} performed LBA observations at 2.3 GHz with slightly shorter baselines detecting TH2 with a flux density of $\sim6$ mJy, a source size of $80 \times 60$ mas ($1.6 \times 1.2$ pc), and a brightness temperature of $\sim10^5$ K. The spectrum of the source between 5-15 GHz is relatively flat ($\alpha \sim -0.47$) and can be best fit with a free-free absorbed-power-law. All these characteristics are not particular unusual in the radio regime and can be explained in terms of either an accreting massive object similar to other synchrotron sources in the centers of active galaxies \citep{hummel84, turner85, gallimore04}, 
or a very compact SNR \citep{ulvestad97}. 

The apparent lack of detected source variability at radio wavelengths 
over 8 year timescales implies that if TH2 is a SNR it is likely not expanding \citep{ulvestad97}. 
Recently, \citet{brunthaler09} reported signs of possible variability of TH2. 
However, as pointed out by these authors, a comparison of different values of flux densities in the literature is difficult, first due to potential contamination by the extended diffuse emission from the galaxy, and second because of the different resolutions used in the observations. This uncertainty can be seen in the comparison of the flux densities from VLA A and B configuration data in \citet{ulvestad97} and in Table 1 and Figure 5 of \citet{lenc06}. The different flux density measurements for TH2 have originated confusion on the nature of its radio spectrum, which has been argued to be flat \citep{turner85}.  


\citet{brunthaler09} question the presence of a possible LLAGN based on the non-detection of TH2 on mas scales with an upper limit of 1 mJy at 23 GHz. In their analysis, they suggest that TH2 may be 
a Gigahertz Peaked Spectrum (GPS) source due to an apparent turnover in its spectrum between 4.8 and 8 GHz.
However, as they pointed out, the GPS nature of TH2 is dubious as GPS sources having a turnover point between 5-10 GHz normally have sizes around 20 pc, $\sim 10$ times larger than the upper limit of 2 pc predicted for TH2, and are several orders of magnitudes more powerful than TH2. In addition, the lack of any optical or IR counterpart is inconsistent with the GPS hypothesis \citep{snellen08}.  



In summary, TH2 is the brightest radio source (a factor $>3$ compared to other sources) detected by the VLA (beam sizes $>100$ mas) in the central $150\times 150$ pc of NGC253. 
However, it does not have any infrared, optical or X-ray counterpart, which argues against an AGN or SNR nature. Regarding the last possibility, \citet{fernandez09} show that a scaled version of the Crab or N40 SNRs to TH2 radio emission level should produce an IR/optical detectable counterpart, which is not the case. 
If indeed associated with the kinematic center of NGC253, 
these properties make TH2 similar to the center of our Galaxy, SgrA$^*$. A dormant BH like SgrA$^*$ at the distance of NGC253 (3.95 Mpc) would have flux densities at radio (23 GHz) and IR (flares at 3.8 $\mu$m) wavelengths of $\sim 3.6\, \mu$Jy and $\sim 0.02\, \mu$Jy, respectively, making the source undetectable. 
The existence of a scaled version of SgrA$^*$ in NGC253 was proposed by \citet{fernandez09} based on the IR/radio ratio of TH2, measured with VLA 15 GHz and NACO $L-$band images of comparable resolution ($\sim 200$ mas, 4 pc), which resembles that of SgrA$^*$ rather than an AGN. 


\subsection{X-1 as candidate for galactic nucleus}\label{hx1}

\textsl{Chandra} observations reveal a heavily absorbed source of hard X-rays 
embedded within the $3\sigma$ error circle of the kinematic center (see Fig.~\ref{fig6}). The hard X-ray emission map from \citet{weaver02} 
shows a partly extended component along the nuclear
ridge confined to an encircled area of r$\sim 2.5^{\prime\prime}$. 
At these scales, it is not clear whether
or not the peak in hard X-rays is pointlike, 
but it has been argued that it is coincident with TH2. 
We reprocessed \textsl{Chandra} data 
and found, with an accurate astrometry of $\sim 0.5^{\prime\prime}$ rms, a central hard X-ray peak (X-1) that is indeed point-like but which lies 
$\sim 1.1^{\prime\prime}$ SW from TH2. It is evident from the similarity of the morphologies between the two hard X-ray maps 
that X-1 corresponds to the strongest source in the 4.5-8 keV continuum map of \citet{weaver02}. 

\begin{figure}
\epsscale{.5}
\plotone{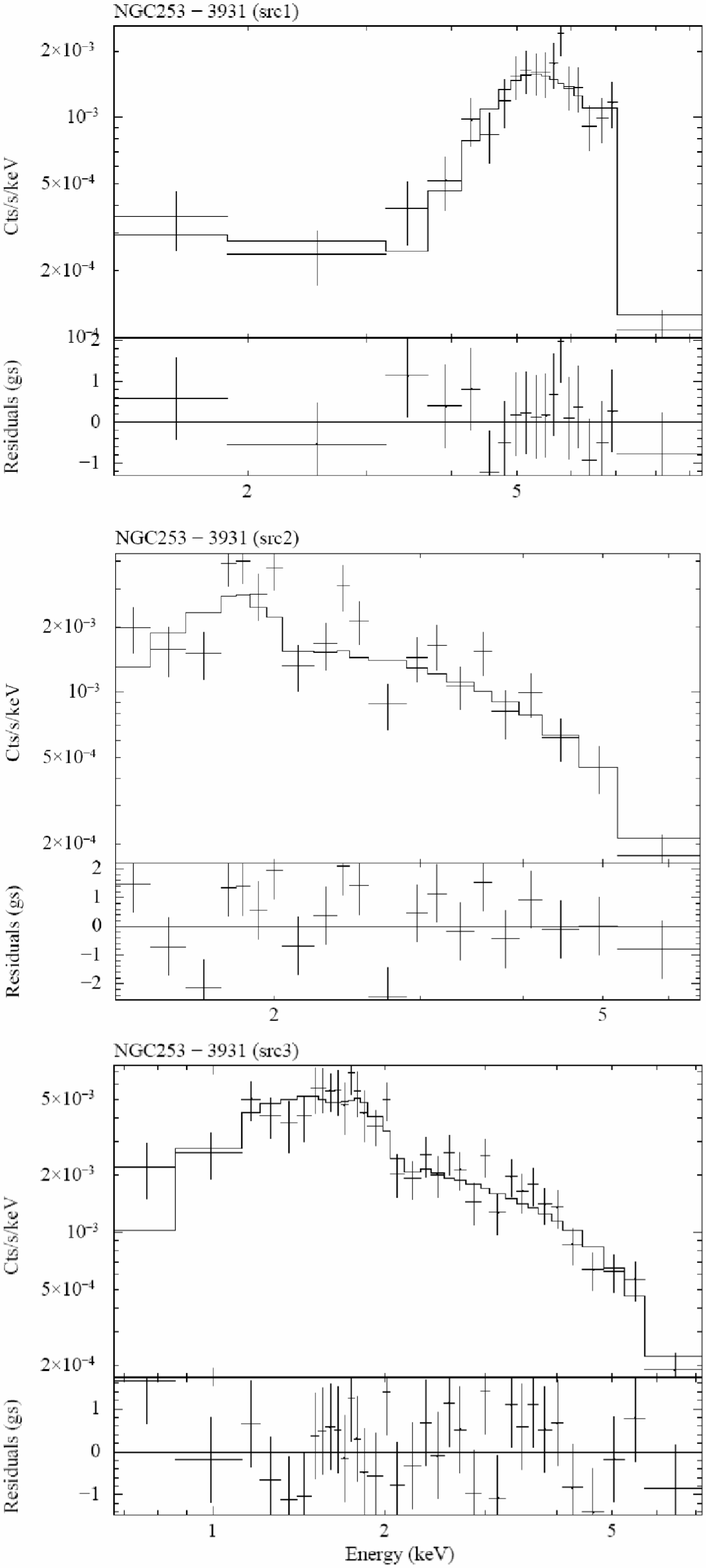}
\caption{Spectral fits to the hard X-ray sources X-1 (top), X-2 (middle), and X-3 (bottom). See Table 1 for model parameters.\label{fig3}}
\end{figure}

\begin{figure}
\epsscale{.99}
\plotone{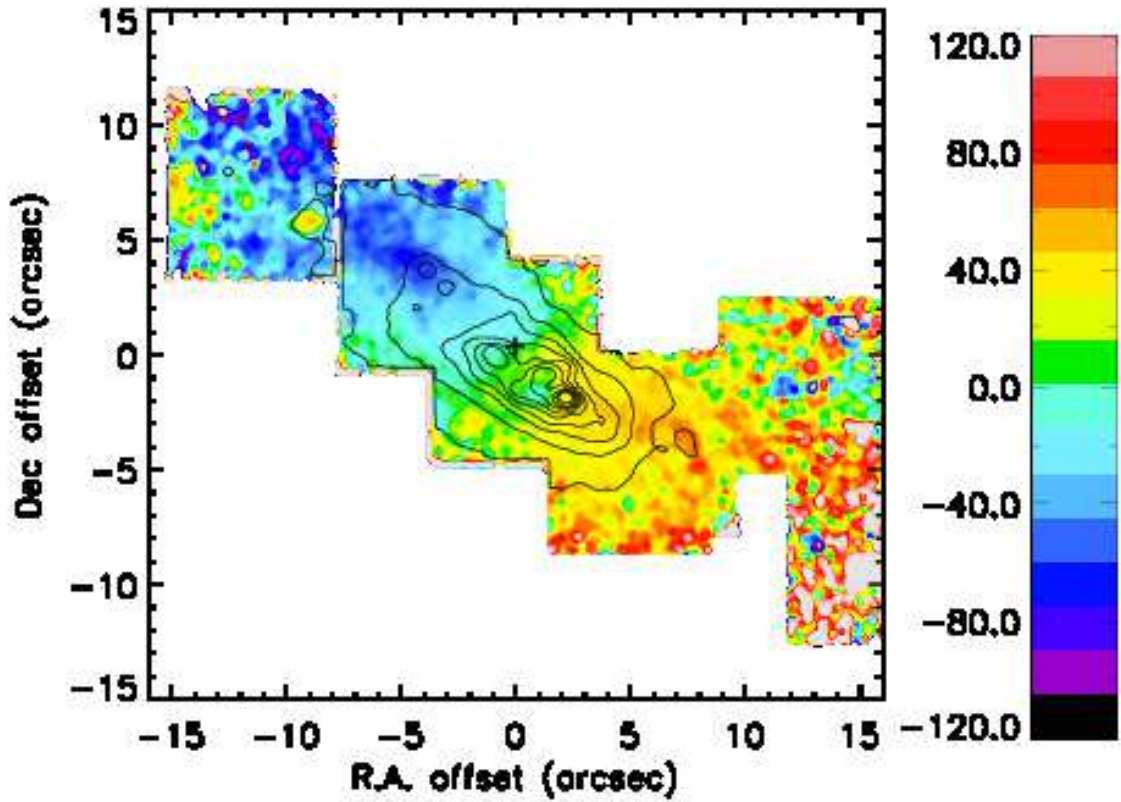}
\caption{Two dimensional velocity field of the stars in the central $30\arcsec \times 30\arcsec$ of NGC253. The velocity map is centered at the position of TH2 (black cross) and the black contours delinate the $K-$band continuum emission. North is up and East is left.\label{fig4}}
\end{figure}

\begin{figure}
\epsscale{.99}
\plotone{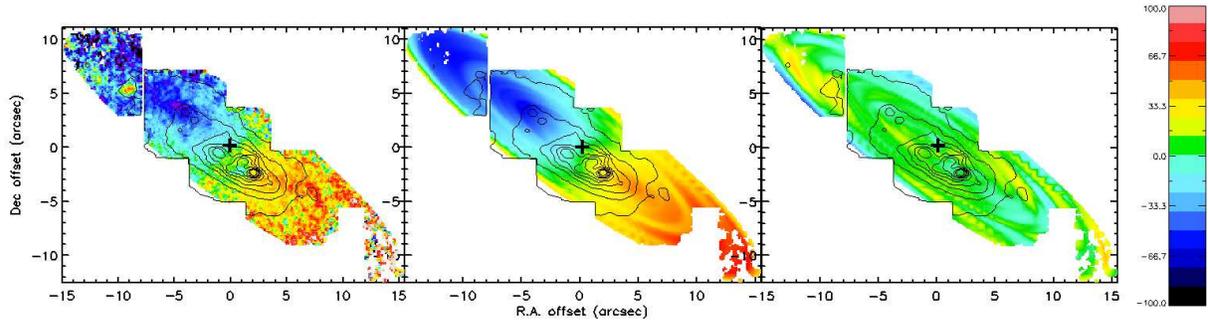}
\caption{Maps showing the data and the best-fit kinematic model used to determine the kinematic center of the galaxy. The maps are centered at the position of TH2 (black cross) and the contours delinate the $K-$band continuum emission. \textit{Left:} Stellar velocity field used for the modeling, \textit{Middle:} Best-fit kinematic model of a rotating disc, \textit{Right:} Residuals (data-model) to the fit.\label{fig5}}
\end{figure}

\begin{figure}
\epsscale{.99}
\plotone{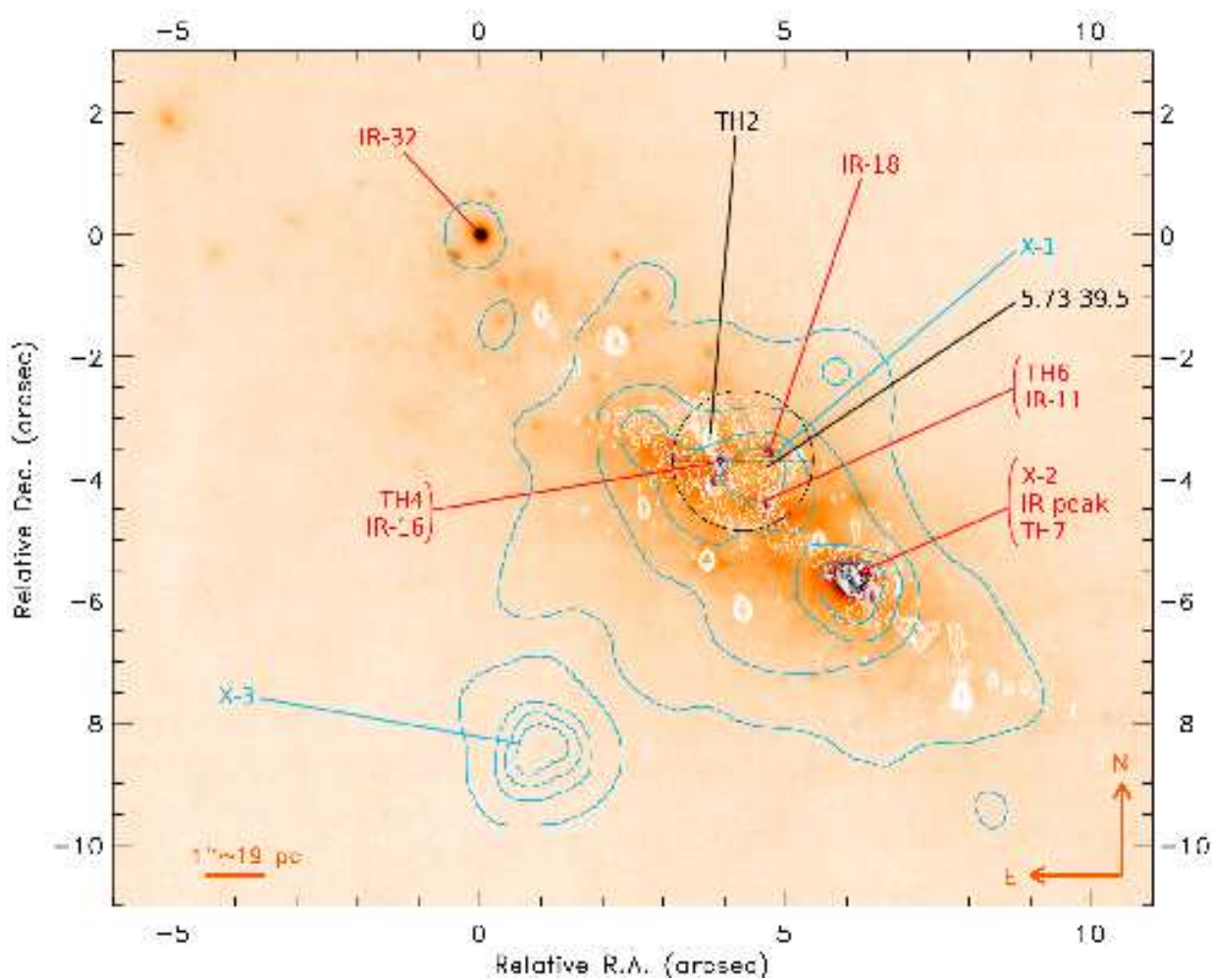}
\caption{Nuclear region of NGC\,253 at infrared NACO L-band (3.8 $\mu$m, colour scale), radio VLA-A 15 GHz (white contours), and hard X-ray 2-10 keV (blue contours) wavelengths. Relevant radio, infrared and X-ray sources are indicated in the Figure. We use the nomenclature from \citet{ulvestad97} for the radio sources  ``TH$n$'' (except for source 5.73--39.5). Infrared sources are labeled as ``IR-$n$'' following the enumeration of \citet{fernandez09} (except for the IR peak), and the X-ray sources are marked as ``X-$n$''. The $3\sigma$ confidence interval of the stellar kinematic center location is approximated by a crossed circle of radius $r=1.2 \arcsec$.  
\label{fig6}}
\end{figure}

\begin{figure}
\epsscale{.99}
\plotone{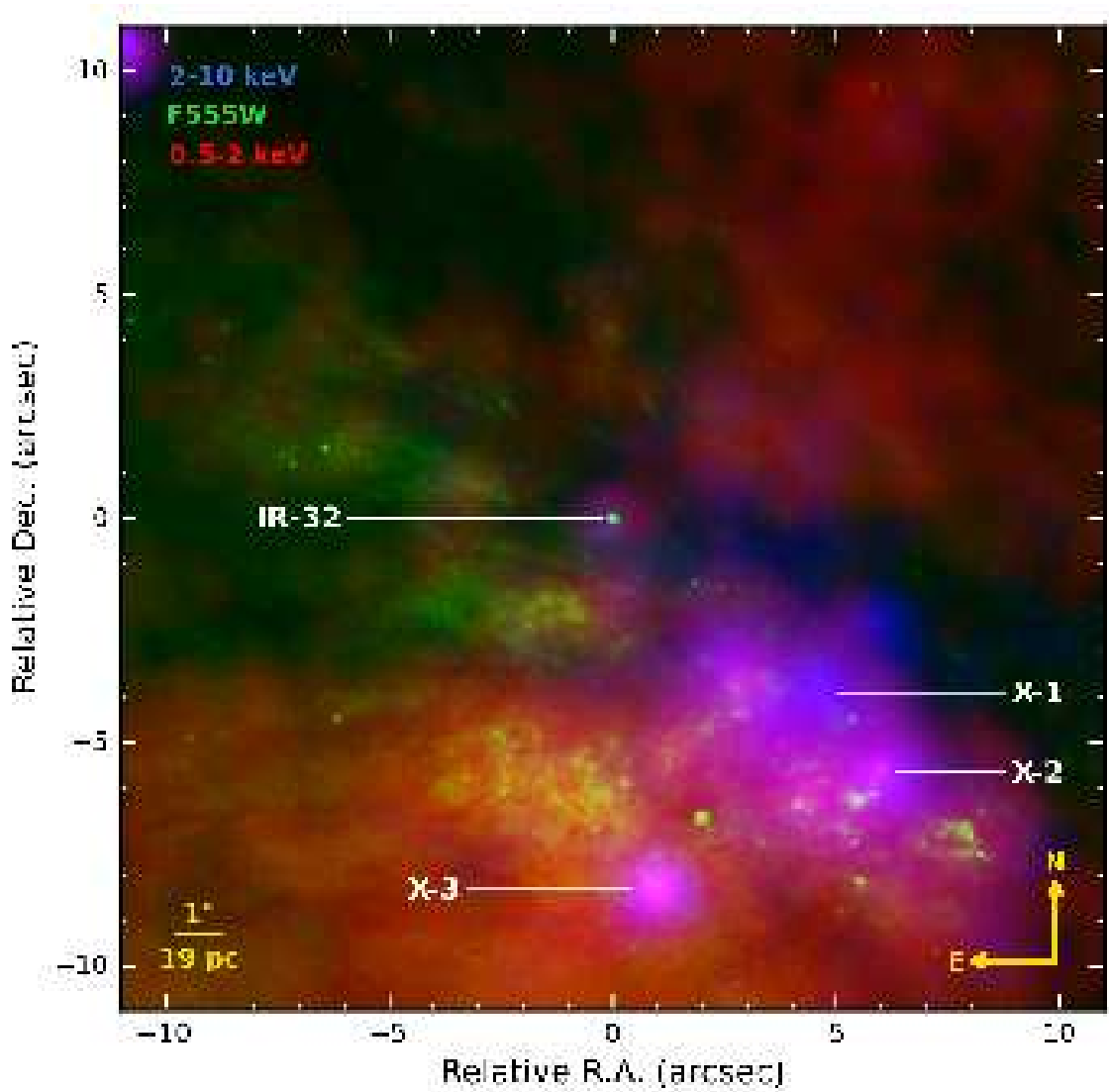}
\caption{Three-color composite image of \textit{Chandra} soft X-ray (red), hard X-ray (blue) and \textit{HST} $V-$band (green) emission in the central $20\arcsec \times 20\arcsec$ region of NGC253. The image is centered at the position of source IR-32, which is an outstanding feature at IR, optical and X-ray wavelengths, and a reference object in the alignment. 
\label{fig7}}
\end{figure}

While the X-ray images from the two studies are very consistent, the X-ray spectrum of X-1 obtained by \citet{weaver02} is very different to the one measured in this work (see Figure 3 of Weaver et al. 2002 and our Fig.~\ref{fig3}).
We believe that this discrepancy is caused by the different aperture sizes used in the extraction of the spectra. In their work, a nuclear spectrum was extracted from within a circular region of diameter $5\arcsec$ centered $1\arcsec$ to the northeast of TH2 in order to avoid contamination from X-2. However, this aperture is with high probability contaminated by the diffuse 0.5--8 keV X-ray emission surrounding X-1 (Fig.~\ref{fig2}).

The X-ray spectrum of X-1 presented in Fig.~\ref{fig3} is best fitted with two absorbed power-laws with $N_{H1} = 4 \times 10^{22}$ and $N_{H2} = 7.5 \times 10^{23}$ cm$^{-2}$ (see Sect.~\ref{xspectra} and Table~\ref{table1}). While $N_{H1}$ is consistent with the extinction of the nuclear starburst region measured with near-IR observations \citep{fernandez09}, $N_{H2}$ is particularly high, indicating that the source is heavily absorbed. 
This is supported by the lack of emission in the soft X-ray band. We hardly detected photons with energies below 2 keV resulting in a very low $F_{0.5-2 keV}/F_{2-10 keV}$ ratio of $3\times 10^{-3}$. 
X-ray sources with low $F_{0.5-2 keV}/F_{2-10 keV}$ and high $N_{H}$ are most often regarded as hidden AGN 
\citep{ueda07, winter09}. 
We estimate the intrinsic 2--10 keV luminosity of X-1 to be $1 \times 10^{40}$ ergs s$^{-1}$, 
consistent with the lower limit proposed by \citep{weaver02} of $2\times 10^{39}$ ergs s$^{-1}$. The intensity of the X-ray emission places the source in the ``ultraluminous X-ray source'' category. 
Given its location, coincident with the kinematic center and embedded in the nuclear starburst, chances are high that the ultraluminous source is a central massive object, either a weakly accreting massive BH (a LLAGN, Terashima et al. 2002; Gonz\'alez-Mart\'in et al 2009) or an evolved NSC brightening up in X-rays most probably due to young SNe and/or SNRs, similar to the luminous X-ray source near the center of M82 \citep{stevens99}.
We examine these two possibilities below. 



Within the achieved $0.5\arcsec$ alignment error between radio, IR and X-ray images (Sect.~\ref{align}), there are two SSCs that could be associated with X-1: IR-11 and IR-18 (Fig.~\ref{fig6}). The two SSCs are spatially resolved in $L-$band, and are very similar in size (FWHM $\sim 3-5$ pc), flux density ($\sim 2-2.5$ mJy in $L$-band) and shape of their IR SEDs \citep{fernandez09}. While IR-11 has an individual counterpart at radio wavelengths (TH6 in Ulvestad \& Antonucci 1997), there is no discrete radio source associated with IR-18 (see Fig.~\ref{fig6}). 
In the radio regime, TH6 is the second strongest flat-spectrum source in NGC253 ($\alpha \sim -0.01$, Lenc \& Tingay 2006). Notice that it is not clear whether the spectrum of TH2 is flat or not (see Sect.~\ref{th2}). 
It appears to be resolved in VLA observations at 23 and 15 GHz (sizes of 4.5 $\times$ 2.4 and 4 $\times$ 1.8 pc, respectively), 
but unresolved at 8.4, 5 and 2.3 GHz. It was not detected at 1.4 GHz with the LBA probably due to free-free absorption as TH2 \citep{tingay04}. Interestingly, this source shows a position offset in LBA observations at 2.3 GHz that is not as consistent as the other sources when compared against VLA positions \citep{lenc06}. These facts suggest that TH6 may have structure that varies significantly with frequency. In addition, it is associated with Radio Recombination Lines (RRL) emission that is consistent with the free-free absorbed spectra of HII regions. 
As in young stellar clusters, hard X-ray radiation 
is mainly produced by young SNe and/or SNRs \citep{stevens99,vanbever00}, the SSC IR-11/radio-TH6 is the most likely counterpart of X-1, presenting the expected simultaneous contribution in radio, IR and X-ray from young SNe and/or SNRs evolving in the SSC \citep{fernandez09}. The fact that IR-18 has no radio counterpart makes more difficult to justify its association with X-1. 
There is an additional radio source, 5.73--39.5 in \citet{lenc06}, also compatible with the location of X-1 (Fig.~\ref{fig6}). However, its flux density has been measured only at two frequencies: 8.4 and 23 GHz. In these measurements 5.73--39.5 is four times weaker than TH6, and has no IR counterpart, all together making dubious any interpretation. 

A second alternative is the association of X-1 with a hidden LLAGN on the basis of the hardness of its X-ray spectrum with an absorbing column density of almost $10^{24}$ cm$^{-2}$. No other manifestation of the active nucleus is found, neither a strong radio nor a near-infrared point source is located at the position of X-1. The absence of any signature other than in the hard X-rays implies that the AGN is non-standard and similar to the hidden AGN detected in NGC4945, but $\sim100$ times less luminous \citep{marconi00}.  

In summary, X-1 appears to be a heavily absorbed object ($N_{H} > 10^{23}$ cm$^{-2}$) only detected at energies $>2$ keV ($L_{2-10 keV} \sim 10^{40}$ ergs s$^{-1}$). If indeed associated with the kinematic center of the galaxy, the simplest explanation accounting for these properties is that X-1 is a hidden LLAGN, similar to the one observed in NGC4945. 
Alternatively, within the \textrm{Chandra} error circle of $r\sim0.5\arcsec$, X-1 could be associated with two SSCs (IR-18 and IR-11/radio-TH6) and one resolved radio source (5.73--39.5). Among these three possibilities, the SSC IR-11/radio-TH6 is the most likely counterpart to X-1. In this case, the hard X-ray emission would be associated to processes occuring in the SSC which could produce the intrinsic 2--10 keV luminosity of $10^{40}$ ergs s$^{-1}$, being the most plausible explanation young SNe and/or SNRs evolving in this massive cluster. The nature of the kinematic center of NGC253 would then be a NSC of $10^5-10^6$ M$_\odot$ with properties similar to other clusters in the nuclear region \citep{fernandez09, mueller09}. 



X-1 might alternatively be a background X-ray source. In this case, the inferred luminosity is only
a lower limit and a background AGN could well be invoked to explain
the X–ray power of the source. Although we cannot completely rule out this scenario, it seems very unlikely that there exists a background AGN which is precisely coincident with the kinematic center of NGC253. In addition, as it was discussed above, X-1 may have radio and/or IR counterparts which are believed to belong to the galaxy.  

\section{Conclusions} \label{conclusions}

SINFONI near-infrared integral field spectroscopy combined with \textit{Chandra} imaging and spectroscopy have been used in this work to probe the galactic nucleus of NGC253. Thanks to our high S/N integral field data we were able to extract 2D stellar kinematics in the nuclear region of NGC253 and estimate the kinematic center of the galaxy with a very good accuracy. Our main results and conclusions can be summarized as follows:

\begin{itemize}
	\item  The stellar kinematic center is located in between the strongest compact radio source (TH2) and the IR photometric center, or IR peak. It lies $\sim 0.7 \arcsec$ SW from TH2 and $\sim 2.6 \arcsec$ NE from the IR peak.
	\item The estimated confidence interval of our kinematic modeling ($r\sim 1.2\arcsec$) clearly discards the IR peak as kinematic center, but includes the position of TH2. 
	\item Newly processed \textsl{Chandra} images reveal a central point-like hard X-ray source (X-1) lying $\sim0.4 \arcsec$ SW from the kinematic center. 
	\item Very accurate alignment between radio, IR and X-ray sources in the nuclear region shows that TH2 and X-1 are not associated with each other, being separated by $\sim 1.1 \arcsec$. As the two could be a manifestation of nuclear activity, and our current estimate of the kinematic center location is consistent with both sources, we consider the two as possible galactic nucleus candidates. 
	\item TH2 is the brightest radio source, with the highest brightness temperature, detected with the VLA in the central region of NGC253. However, it does not have any IR, optical or X-ray counterparts. 
When observed at very high spatial resolution with the VLBI, TH2 is among the weakest sources ($\sim 6$ mJy at 2.3 GHz, Lenc \& Tingay 2006), or it is not detected ($<1$ mJy at 23 GHZ, Brunthaler et al. 2009).
If the kinematic center is associated with TH2, we conclude that the nature of the nucleus of NGC253 resembles that of SgrA$^*$. 
\item X-1 is a heavily absorbed object ($N_{H} > 10^{23}$ cm$^{-2}$) only detected at energies $>2$ keV ($L_{2-10 keV} \sim 10^{40}$ ergs s$^{-1}$). 
If X-1 corresponds to the kinematic center of the galaxy, 
the simplest explanation accounting for these properties is that the nucleus of NGC253 harbors a hidden LLAGN. 
\item Alternatively, X-1 may be associated with one SSC in this region: SSC IR-11/radio-TH6. In this case the kinematic center of NGC253 would then be a $10^5-10^6$ M$_\odot$ spatially resolved (diameter of $\sim5$ pc) NSC in which some of its massive stars are evolving into the SN phase. 
\end{itemize}

An important conclusion from this study is that the nature of dusty galaxies cannot be always characterized by photometric data alone but must be complemented with spectroscopic observations. In particular, we demonstrated the power of integral field spectroscopy for the determination of kinematic centers of galaxies.
Further near-IR integral-field observations with high spatial resoultion would be useful to derive a more accurate position of the kinematic center, measure directly the mass of the putative central massive object, and reveal the presence of either broad recombination lines or forbidden high ionization lines. A detection of any of these would provide definite evidence of the existence of an AGN in this galaxy.



\acknowledgments

We thank the referee for helpful comments that helped improve the manuscript.
FMS acknowledges a Research Fellowship from the Spanish MEC project AYA2007-60235. 
OGM acknowledges support by the EU FP7-REGPOT 206469 and ToK 39965 grants. 



Facilities: \facility{VLT:Yepun(SINFONI)}, \facility{CXO(ACIS)}.

\clearpage



\clearpage

\clearpage

\clearpage




\clearpage

\clearpage

\clearpage

\clearpage

\begin{table}
\begin{center}
\caption{Spectral models for the hard X-ray sources found in the nucleus of NGC253.\label{table1}}
\begin{tabular}{lccccccccc}
\tableline\tableline
Source & Type\tablenotemark{a} & Counts\tablenotemark{b} & $N_{H1}$\tablenotemark{c} & $N_{H2}$\tablenotemark{c} & $\Gamma$ & $F_{0.5-2 keV}$\tablenotemark{d} & $F_{2-10 keV}$\tablenotemark{d} & 
$L_{2-10 keV}$\tablenotemark{e} & $\chi^2$/dof \\
\tableline
X-1 & 2po & 393 & $4.0_{-1.9}^{+3.7}$ & $75_{-12}^{+24}$ & $3.6_{-1.2}^{+0.8}$ & $-15.0_{-1.3}^{+0.1}$ & $-12.64_{-0.31}^{+0.03}$ & 
$40.03_{-1.0}^{+0.02}$ & $15.0/15$\\
X-2 & po & 436 & $2.5_{-0.7}^{+0.8}$ & - & $2.3_{-0.3}^{+0.4}$ & 
$-14.2_{-0.3}^{+0.1}$ & $-13.17_{-0.35}^{+0.08}$ & 
$38.18_{-0.33}^{+0.1}$ & $33.1/18$\\
X-3 & po & 803 & $0.7_{-0.2}^{+0.3}$ & - & $1.6_{-0.2}^{+0.2}$ & 
$-13.7_{-0.1}^{+0.1}$ & $-12.98_{-0.08}^{+0.05}$ & 
$38.29_{-0.1}^{+0.06}$ & $32.3/43$\\
\tableline
\end{tabular}


\tablenotetext{a}{Best-fit spectral model, $po$ stands for abs(power-law), $me$ for abs(MEKAL), and 2po for abs(power-law)+abs(power-law).}
\tablenotetext{b}{Number of counts in the range 0.5-10 keV}
\tablenotetext{c}{Hydrogen column densities in units of $10^{22}$ cm$^{-2}$}
\tablenotetext{d}{Observed flux in logarithmic units of ergs cm$^{-2}$ s$^{-1}$}
\tablenotetext{e}{Intrinsic hard X-ray luminosity in logarithmic units of ergs  s$^{-1}$ (assuming $d=3.95$ Mpc)}
\tablecomments{All quoted untertainties are $90\%$ except for the fluxes and luminosities, which are $1\sigma$.}
\end{center}
\end{table}







\begin{thebibliography}{}

\bibitem[Abuter et al.(2005)]{abuter05} Abuter, R., Schreiber, J., Eisenhauer, F., Ott, T., Horrobin, M., \& Gillesen, S. 2005, NewAR, 50, 398 


\bibitem[Anantharamaiah \& Goss(1996)]{anantamariah96} Anantharamaiah, K. R., \& Goss, W. M. 1996, ApJ, 466, L13

\bibitem[Arnaboldi et al.(1995)]{arnaboldi95} Arnaboldi, M., Capaccioli, M., Cappellaro, E., Held, E. V., \& Koribalski, B. 1995, AJ, 110, 199


\bibitem[Bonnet et al.(2004)]{bonnet04} Bonnet, H., et al. 2004, The ESO Messenger, 117, 17

\bibitem[B\"oker et al.(1998)]{boeker98} B\"oker, T., Krabbe, A. \& Storey, J. W. V. 1998, ApJ, 498, 115

\bibitem[B\"oker et al.(2002)]{boeker02} B\"oker, T., Laine, S., van der Marel, R. P., Sarzi, M., Rix, H.-W., Ho, L. C., \& Shields, J. C. 2002, AJ, 123, 1389 

\bibitem[B\"oker et al.(2008)]{boeker08} B\"oker, T., Falc\'on-Barroso, J., Schinnerer, E., Knapen, J. H., \& Ryder, S. 2008, AJ, 135, 479

\bibitem[Brunthaler et al.(2009)]{brunthaler09} Brunthaler, A., Castangia, P., Tarchi, A., Henkel, C., Reid, M. J., Falcke, H., \& Menten, K. M. 2009, A\&A, 497, 103

\bibitem[Canzian, Mundy \& Scoville(1988)]{canzian88} Canzian, B., Mundy, L. G., \& Scoville, N. Z. 1988, ApJ, 333, 157




\bibitem[Das et al.(2001)]{das01} Das, M., Anantharamaiah, K. R., \& Yun, M. S. 2001, ApJ, 549, 896


\bibitem[Davies et al.(2007)]{davies07} Davies, R., M\"uller Sanchez, F., Genzel, R., Tacconi, L.J., Hicks, E., Friedrich, S., \& Sternberg, A. 2007, ApJ, 671, 1388

\bibitem[Ebeling et al.(2006)]{ebeling06} Ebeling, H., White, D. A., \& Rangarajan, F. V. N. 2006, MNRAS, 368, 65

\bibitem[Eisenhauer et al.(2003)]{eisenhau03} Eisenhauer, F., et al. 2003, in Instrument Design and Performance for Optical/Infrared Ground-based Tele-
scopes, eds. Masanori I., Moorwood A., Proc. SPIE, 4841, 1548

\bibitem[Engelbracht et al.(1998)]{engelbracht98} Engelbracht, C. W., Rieke, M. J., Rieke, G. H., Kelly, D. M., \& Achtermann, J. M. 1998, ApJ, 505, 639

\bibitem[Fern\'andez-Ontiveros et al.(2009)]{fernandez09} Fern\'andez-Ontiveros, J. A., Prieto, M. A., \& Acosta-Pulido, J. A. 2009, MNRAS, 392, L16

\bibitem[Filippenko \& Sargent(1989)]{filippenko89} Filippenko, A. V., \& Sargent, W.L.W. 1989, ApJ, 342, L11 

\bibitem[Forbes et al.(2000)]{forbes00} Forbes, D. A., Polehampton, E., Stevens, I. R., Brodie, J. P., \& Ward, M. J. 2000, MNRAS, 312, 689

\bibitem[Galliano et al.(2005)]{galliano05} Galliano, E., Alloin, D., Pantin, E., Lagage, P. O., \& Marco, O. 2005, A\&A, 438, 803

\bibitem[Gallimore et al.(2004)]{gallimore04} Gallimore, J. F., Baum, S. A., \& O'Dea, C. P. 2004, ApJ, 613, 794

\bibitem[Gonz\'alez-Mart\'in et al.(2009)]{gonzalez09} Gonz\'alez-Mart\'in, O., Masegosa, J., M\'arquez, I., \& Guainazzi, M. 2009, ApJ, 704, 1570

\bibitem[Graham(2008)]{graham08} Graham, A. W. 2008, PASA, 25, 167



\bibitem[Heckman(1998)]{heckman98} Heckman, T. M. 1998, in ASP Conf. Ser. 148, Origins, ed. C. E. Woodward, J. M. Shull, \& H. A. Thronson, Jr. (San Francisco: ASP), 127

\bibitem[Hummel et al.(1984)]{hummel84} Hummel, E., van der Hulst, J. M., \& Dickey, J. M. 1984, A\&A, 134, 207

\bibitem[Karachentsev et al.(2003)]{kara03} Karachentsev, I.~D., et al. 2003, A\&A, 438, 803

\bibitem[Keto et al.(1999)]{keto99} Keto, E., Hora, J. L., Fazio, G. G., Hoffmann, W. \& Deutsch, L. 1999, ApJ, 518, 183


\bibitem[Krajnovi\'c et al.(2006)]{krajnovic06} Krajnovi\'c, D., Cappellari, M., de Zeeuw, P. T., \& Copin, Y. 2006, MNRAS, 366, 787

\bibitem[Krajnovi\'c et al.(2007)]{krajnovic07} Krajnovi\'c, D., Sharp, R., \& Thatte, N. 2007, MNRAS, 374, 385 


\bibitem[Lampton et al.(1976)]{lampton76} Lampton, M., Margon, B., \& Bowyer, S. 1976, ApJ, 208, 177


\bibitem[Lehnert \& Heckman(1996)]{lehnert96} Lehnert, M. D., \& Heckman, T. M. 1996, ApJ, 462, 651

\bibitem[Lenc \& Tingay(2006)]{lenc06} Lenc, E., \& Tingay, S.~J. 2006, AJ, 132, 1333

\bibitem[Marconi et al.(2000)]{marconi00} Marconi, A., Oliva, E., van der Werf, P. P., Maiolino, R., Schreier, E. J., Macchetto, F., \& Moorwood, A.F.M. 2000, A\&A, 357, 24


\bibitem[Matsumoto et al.(2001)]{matsumoto01} Matsumoto, H., Tsuru, T. G., Koyama, K., Awaki, H., Canizares, C. R., Kawai, N., Matsushita, S., \& Kawabe, R. 2001, ApJ, 547, L25

\bibitem[Mauersberger et al.(1996)]{mauersberger96} Mauersberger, R., Henkel, C., Wielebinski, R., Wiklind, T., \& Reuter, H.-P. 1996, A\&A, 305, 421


\bibitem[Mohan et al.(2002)]{mohan02} Mohan, N. R., Anantharamaiah, K. R., \& Goss, W. M. 2002, ApJ, 574, 701


\bibitem[M\"uller-S\'anchez et al.(2010b)]{mueller09}  M\"uller-S\'anchez, F., et al. 2010b, in preparation

\bibitem[Mu\~noz-Tu\~non et al. (1993)]{munoz93} Mu\~noz-Tu\~non, C., Vilchez, J. M., \& Casta\~neda, H. 1993, A\&A, 278, 364

\bibitem[Nakai et al.(1995)]{nakai95} Nakai, N., Inoue, M., Miyazawa, K., Miyoshi, M., \& Hall, P. 1995, PASJ, 47, 771

\bibitem[Nelder \& Mead(1965)]{nelder65} Nelder, J.A., \& Mead, R. 1965, Computer Journal, 7, 308 


\bibitem[Paglione et al.(1995)]{paglione95} Paglione, T. A. D., Tosaki, T., \& Jackson, J. M. 1995, ApJ, 454, L117

\bibitem[Paglione et al.(2004)]{paglione04} Paglione, T. A. D., Yam, O., Tosaki, T., \& Jackson, J. M. 2004, 611, 835


\bibitem[Pence(1980)]{pence80} Pence, W. D. 1980, ApJ, 239, 54

\bibitem[Peng et al.(1996)]{peng96} Peng, R., Zhou, S., Whiteoak, J. B., Lo, K. Y., \& Sutton, E. C. 1996, ApJ, 470, 821


\bibitem[Prada et al.(1998)]{prada96} Prada, F., Guti\'errez, C. M., \& McKeith, C. D. 1998, ApJ, 495, 765


\bibitem[Richstone et al.(1998)]{richstone98} Richstone, D., et al. 1998, Nature, 395, A14

\bibitem[Sadler et al.(1995)]{sadler95} Sadler, E. M., Slee, O. B., Reynolds, J. E., \& Roy, A. L. 1995, MNRAS, 276, 1373


\bibitem[Sams et al.(1994)]{sams94} Sams III, B. J., Genzel, R., Eckart, A., Tacconi-Garman, L., \& Hofmann, R. 1994, ApJ, 430, L33


\bibitem[Seth et al.(2010)]{seth10} Seth, A., et al. 2010, ApJ, 714, 713

\bibitem[Shapiro et al.(2008)]{shapiro08} Shapiro, K., et al. 2008, ApJ, 682, 231 

\bibitem[Shapley et al.(2003)]{shapley03} Shapley, A., et al. 2003, ApJ, 588, 65S

\bibitem[Snellen et al.(1998)]{snellen08} Snellen, I. A. G., et al. 1998, MNRAS, 301, 985

\bibitem[Stevens et al.(1999)]{stevens99} Stevens, I. R., Strickland, D. K., \& Wills, K. A. 1999, MNRAS, 308, L23

\bibitem[Strickland et al.(2000)]{strickland00} Strickland, D. K., Heckman, T. M., Weaver, K. A., \& Dahlem, M. 2000, AJ, 120, 2965


\bibitem[Summers et al.(2004)]{summers04} Summers, L. K., Stevens, I. R., Strickland, D. K. \& Heckman, T. M. 2004, MNRAS, 351, 1


\bibitem[Tacconi-Garman et al.(2005)]{tacconi05} Tacconi-Garman, L. E., Sturm, E., Lehnert, M., Lutz, D., Davies, R. I., \& Moorwood, A. F. M. 2005, A\&A, 432, 91

\bibitem[Tacconi-Garman(2006)]{tacconi06} Tacconi-Garman, L. E. 2006, NewAR, 49, 569

\bibitem[Terashima et al.(2002)]{terashima02} Terashima, Y., Iyomoto, N., Ho, L. C., \& Ptak, A. F. 2002, ApJS, 139,1 

\bibitem[Tingay(2004)]{tingay04} Tingay, S. J. 2004, AJ, 127, 10

\bibitem[Trachternach et al.(2008)]{schoenmakers99} Trachternach, C., de Blok, W. J. G., Walter, F., Brinks, E., \& Kennicutt, R. C. 2008, AJ, 136, 2720

\bibitem[Turner \& Ho(1985)]{turner85} Turner, J.~L., \& Ho, P.~T.~P. 1985, ApJL, 299, L77

\bibitem[Ueda et al. (2007)]{ueda07} Ueda, Y., Eguchi, S., Terashima, Y., Mushotzky, R., Tueller, J., Markwardt, C., Gehrels, N.,
Hashimoto, Y., \& Potter, S. 2007, ApJ, 664, L79

\bibitem[Ulrich(1978)]{ulrich78} Ulrich, M. H. 1978, ApJ, 219, 424

\bibitem[Ulvestad \& Antonucci(1997)]{ulvestad97} Ulvestad, J.~S., \& Antonucci, R.~R.~J. 1997, ApJ, 488, 621

\bibitem[Van Bever \& Vanbeveren(2000)]{vanbever00} Van Bever, J., \& Vanbeveren, D. 2000, A\&A, 358, 462

\bibitem[van der Marel et al.(1998)]{vandermarel98} van der Marel, R. P., Cretton, N., de Zeeuw, P. T., \& Rix, H.-W. 1998, ApJ, 493, 613


\bibitem[Weaver et al.(2002)]{weaver02} Weaver, K. A., Heckman, T. M., Strickland, D. K., \& Dahlem, M. 2002, ApJL, 576, L19

\bibitem[Wills et al.(1999)]{wills99} Wills, K. A., Pedlar, A., Muxlow, T.W.B., \& Stevens, I. R. 1999, MNRAS, 305, 680

\bibitem[Winter et al.(2009)]{winter09} Winter, L. M., Mushotzky, R. F., Terashima, Y., \& Ueda, Y. 2009, ApJ, 701, 1644


\end{thebibliography}
\end{document}